\documentclass{article}
\usepackage{spconf,graphicx}
\usepackage{amsmath}
\usepackage{amsfonts}
\usepackage[noend]{algpseudocode}
\usepackage{booktabs}
\usepackage{algorithm}      
\usepackage{algpseudocode}  
\usepackage{url}
\usepackage{hyperref}
\usepackage{multirow}
\usepackage{pifont}
\usepackage{ifthen}
\usepackage[
   backend=biber,
   style=ieee,
   citestyle=numeric-comp,
   maxbibnames=3,
   maxcitenames=3,
   doi=false,isbn=false,url=false,eprint=false
]{biblatex}
\DeclareSourcemap{
  \maps[datatype=bibtex, overwrite=true]{
    \map{
      \step[fieldsource=booktitle, match=\regexp{.*Interspeech.*}, replace={Proc. Interspeech}]
      \step[fieldsource=journal, match=\regexp{.*INTERSPEECH.*}, replace={Proc. Interspeech}]
      \step[fieldsource=booktitle, match=\regexp{.*ICASSP.*}, replace={Proc. ICASSP}]
      \step[fieldsource=booktitle, match=\regexp{.*icassp_inpress.*}, replace={Proc. ICASSP (in press)}]
      \step[fieldsource=booktitle, match=\regexp{.*Acoustics,.*Speech.*and.*Signal.*Processing.*}, replace={Proc. ICASSP}]
      \step[fieldsource=booktitle, match=\regexp{.*International.*Conference.*on.*Learning.*Representations.*}, replace={Proc. ICLR}]
      \step[fieldsource=booktitle, match=\regexp{.*International.*Conference.*on.*Computational.*Linguistics.*}, replace={Proc. COLING}]
      \step[fieldsource=booktitle, match=\regexp{.*SIGdial.*Meeting.*on.*Discourse.*and.*Dialogue.*}, replace={Proc. SIGDIAL}]
      \step[fieldsource=booktitle, match=\regexp{.*International.*Conference.*on.*Machine.*Learning.*}, replace={Proc. ICML}]
      \step[fieldsource=booktitle, match=\regexp{.*North.*American.*Chapter.*of.*the.*Association.*for.*Computational.*Linguistics:.*Human.*Language.*Technologies.*}, replace={Proc. NAACL}]
      \step[fieldsource=booktitle, match=\regexp{.*Empirical.*Methods.*in.*Natural.*Language.*Processing.*}, replace={Proc. EMNLP}]
      \step[fieldsource=booktitle, match=\regexp{.*Association.*for.*Computational.*Linguistics.*}, replace={Proc. ACL}]
      \step[fieldsource=booktitle, match=\regexp{.*Automatic.*Speech.*Recognition.*and.*Understanding.*}, replace={Proc. ASRU}]
      \step[fieldsource=booktitle, match=\regexp{.*Spoken.*Language.*Technology.*}, replace={Proc. SLT}]
      \step[fieldsource=booktitle, match=\regexp{.*Speech.*Synthesis.*Workshop.*}, replace={Proc. SSW}]
      \step[fieldsource=booktitle, match=\regexp{.*workshop.*on.*speech.*synthesis.*}, replace={Proc. SSW}]
      \step[fieldsource=booktitle, match=\regexp{.*Advances.*in.*neural.*information.*processing.*}, replace={Proc. NeurIPS}]
      \step[fieldsource=booktitle, match=\regexp{.*Advances.*in.*Neural.*Information.*Processing.*}, replace={Proc. NeurIPS}]
      \step[fieldsource=booktitle, match=\regexp{.*Workshop.*on.*Applications.*of.*Signal.*Processing.*to.*Audio.*and.*Acoustics.*}, replace={Proc. WASPAA}]
      \step[fieldsource=publisher, match=\regexp{.+}, replace={{}}]
      \step[fieldsource=month, match=\regexp{.+}, replace={{}}]
      \step[fieldsource=location, match=\regexp{.+}, replace={{}}]
      \step[fieldsource=address, match=\regexp{.+}, replace={{}}]
      \step[fieldsource=organization, match=\regexp{.+}, replace={{}}]
    }
  }
}

\defbibheading{bibliography}[\refname]{}

\title{SPEECH COLLAGE: CODE-SWITCHED AUDIO GENERATION
By COLLAGING MONOLINGUAL CORPORA}
\name{
\begin{tabular}{c}
Amir Hussein\thanks{$^\dagger$Both authors contributed equally to this research.}~$^{\dagger 1}$, Dorsa Zeinali~$^{\dagger 2}$, Ondřej Klejch$^3$, Matthew Wiesner$^1$,  Brian Yan$^4$,
\\ Shammur Chowdhury$^5$, Ahmed Ali$^5$, Shinji Watanabe $^4$, Sanjeev Khudanpur$^1$
\end{tabular}
}
\address{$^1$Johns Hopkins University, USA, $^2$Northeastern University, USA,$^3$ University of Edinburgh, UK, \\$^4$Carnegie Mellon University, USA, $^5$ Qatar Computing Research Institute, Doha}
\begin{document}

\ninept
\maketitle
\begin{abstract}
Designing effective automatic speech recognition (ASR) systems for Code-Switching (CS) often depends on the availability of the transcribed CS resources. To address data scarcity, this paper introduces \emph{Speech Collage}, a method that synthesizes CS data from monolingual corpora by splicing audio segments. We further improve the smoothness quality of audio generation using an overlap-add approach. We investigate the impact of generated data on speech recognition in two scenarios: using in-domain CS text and a zero-shot approach with synthesized CS text. Empirical results highlight up to 34.4\% and 16.2\% relative reductions in Mixed-Error Rate and Word-Error Rate for in-domain and zero-shot scenarios, respectively. Lastly, we demonstrate that CS augmentation bolsters the model's code-switching inclination and reduces its monolingual bias.
\end{abstract}
\section{Introduction}
\label{sec:intro}
In multilingual societies, code-switching (CS) is integral to
communication, enabling clearer expression and reflecting
cultural nuances \cite{scotton1982possibility,sitaram2019survey}. While CS is prevalent in daily conversations, it is underrepresented in transcribed datasets. This linguistic phenomenon, where speakers interweave languages within a conversation or utterance, poses challenges for voice technologies like automatic speech recognition (ASR). Given the abundance of monolingual data and scarcity of labeled CS speech, there's a pressing need to harness monolingual resources for CS applications. The prime challenge lies in developing robust ASR systems for CS in zero-shot settings where no CS training data is available. 

Several approaches have proposed to build CS ASR directly from monolingual data by utilizing multilingual training \cite{taneja2019exploiting,liu2021code,chuang2020training,shah2020learning,chowdhury2021towards,ali2021arabic}. Further studies advocate for the joint modeling of CS and monolingual ASR, effectively breaking down bilingual tasks into monolingual components \cite{zhou2020multi,yan2022joint,yan2023towards}.
A prominent issue with monolingual training is the model's monolingual bias which impedes seamless language switching \cite{hussein2021balanced}.
To address this issue, several data augmentation strategies have been proposed including textual data augmentation, text-to-speech synthesis, and concatenation-based speech generation. In \cite{hussein2022code} authors proposed a methodology to generate the code-switching text from monolingual text to improved ASR performance with language model rescoring. In \cite{seki2018end, ali2021arabic}, researchers propose merging monolingual utterance to mimic code-switching. However, this strategy tends to primarily capture inter-sentential switches, often sidelining the nuances of intra-sentential CS. On another front, text-to-speech (TTS) based synthetic audio has gained traction for CS data generation \cite{8693044,sharma2020improving,murthy2018effect,peyser2019improving,rosenberg2019speech,wang2020improving,yu2023code}.  Despite its potential, TTS based augmentation suffer from limited speaker variability compared to real data. Consequently, there's a growing interest in using audio segment splicing as augmentation to covers more speaker variations and acoustic environments \cite{lam21b_interspeech,zhao2021addressing}. However in the proposed splicing, speech segments and their corresponding words are randomly selected, and the potential of splicing method in code-switching remains unexplored.

In this paper, we introduce \emph{Speech Collage}\footnote{Visit our repository for audio samples and implementation \url {https://github.com/JSALT2022CodeSwitchingASR/generating-code-switched-audio}}, a data augmentation technique that constructs synthetic code-switched audio from monolingual data. Our method is inspired by traditional concatenation-based speech synthesis techniques \cite{hunt1996unit,wouters2001control}.
We demonstrate the efficacy of \emph{Speech Collage} with two scenarios: a) In-domain CS text: where target-domain CS text is leveraged, and b) Zero-shot CS: where synthesized CS text is used. Our study covers two language pairs: Mandarin-English and Arabic-English. 
Experimental results show substantial improvements \emph{Speech Collage} brings to code-switching ASR for both scenarios. Our contributions include: (i) a novel speaker-agnostic CS data augmentation derived from monolingual resources, (ii) further improving ASR performance with enhanced audio quality in generated data, and (iii) propose a of zero-shot learning framework tailored for CS. 
As an additional contribution, we conduct an ablation study to assess the significance of each component on the final performance. We also perform a modified Code Mixed Index (CMI) analysis to identify where the primary gains achieved through our augmentation method.
\begin{figure*}[hbt!]
\centering
\includegraphics[width=14cm,height=5.7cm]{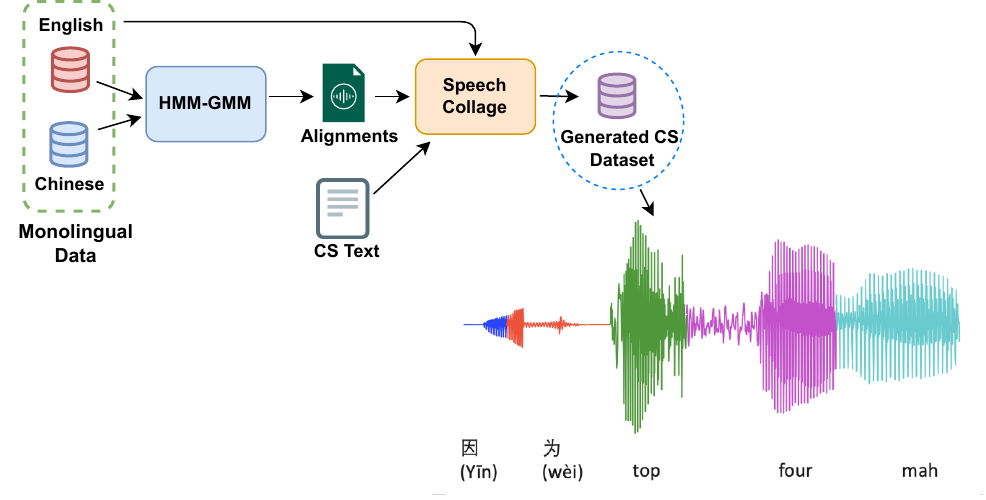}
\caption{High level illustration of the proposed Speech Collage CS generation approach.}
\label{collage-diagram} 
\end{figure*}

\section{Speech Collage}
\label{sec:speech-collage}
We propose a framework designed to splice speech units extracted from monolingual corpora. These units are based on code-switched text, either real or synthesized, as depicted in Figure\ref{collage-diagram}. For the merging process, we select word-units for English and Arabic, and characters for Mandarin. While smaller units, such as phones, offer greater adaptability, they tend to degrade audio quality \cite{khan2016concatenative}. The constructed data form segment splicing, encompasses variations from multiple speakers and diverse acoustic environments. We first obtain the unit alignments with audio from the monolingual data by training standard Hidden Markov Model-Gaussian Mixture Model (HMM-GMM) \footnote{
\url{https://github.com/kaldi-asr/kaldi/tree/master/egs/aishell/s5}
\par
\url{https://github.com/kaldi-asr/kaldi/tree/master/egs/mgb2_arabic/s5}
} using Kaldi ASR toolkit \cite{povey2011kaldi}. Utilizing these alignments, in conjunction with the CS text and monolingual audio, our Speech Collage framework generates the CS audio dataset. In cases where the training data possesses multiple segments for a singular unit, a segment is selected at random. The generated audio quality is further enhanced using the overlap-add technique, energy normalization, and n-gram matching, as detailed below. The audio enhancement and segment splicing were implemented using the Lhotse toolkit \cite{zelasko2021lhotse}.

\subsection{Overlap-add}
\label{subsec:overlap-add}
To enhance the quality of the generated CS audio, we employ the overlap-add with a Hamming window to mitigate discontinuity effects resulting from spliced units. To ensure the segment capture of each unit, we extend the unit-segments by $0.05$ seconds at the start and end of the segment. This extension provides an extra $0.05$ second which is utilized as overlap in overlap-add process.

\subsection{Energy normalization}
\label{subsec:energy-norm}
Additionally, we normalize the synthesized utterance by the average of unit-segments energy to remove artifacts introduced by energy variations between segments. For a speech sequence $X$ of length $T$, $X = \{x_t \in \mathbb{R}| t = 1, \cdots, T \}$, the average energy is calculated as follows: 
\begin{equation}\label{eq1}
X' = \left \{ \frac{x_t}{\sqrt{\frac{1}{T}\sum_t x_t^2}} |t = 1, \cdots, T\right \}
\end{equation}

\subsection{N-gram units}
\label{subsec:n-grams}
To further enhance the quality of the generated CS we explore splicing consecutive units (n-grams), in alignment with selecting longer units in concatenated speech synthesis \cite{black1997automatically} Given a CS sentence our approach starts by matching the largest consecutive unit from monolingual alignments. If a specific n-gram is unavailable, the algorithm backs off to a smaller unit. It's worth noting that in this study, we only experimented with unigrams and bigrams. A detailed description of n-gram Speech Collage implementation is described in Algorithm \ref{alg:speech-collage}. 
\begin{algorithm}[t]
\caption{Speech Collage}\label{alg:speech-collage}
\begin{algorithmic}[1]
\Require Monolingual combined corpus $\mathcal{D}$
\Require Code-switched text $\mathcal{T}_{CS}$
\Require Alignments $\mathcal{A}$ , size of n-grams $n$

\Function{SetupSupervisions}{$\mathcal{A}, n$}
    \For {$w_{1:n}$ with timings $(t^n_s,t^n_e)$ $\in \mathcal{A}$}
        \State cut, $w_{1:i}$ $\gets$ \textsc{getLongestAudioCut}($w_{1:n}$, $(t^n_s,t^n_e)$)  
        \State $\mathcal{D}[w_{1:i}] \gets$ cut
    \EndFor
    \State \Return $\mathcal{D}$
\EndFunction

\Function{NormalizeEnergy}{audioCut}\Comment{Eq \ref{eq6}}
    \State $e \gets$ \textsc{AvgEnergy}(audioCut) 
    \State \Return audioCut $/$ $\sqrt{e}$
\EndFunction

\Function{SampleUnit}{$\mathcal{D}$, $w_{1:i}$}
    \State \Return random cut from $\mathcal{D}[w_{1:i}]$
\EndFunction

\Function{GenerateCollage}{$\mathcal{T}_{CS}$, $\mathcal{A}$, $n$}
    \State $\mathcal{D} \gets$ \textsc{SetupSupervisions}$(\mathcal{A},n)$ \Comment{$n$: size of n-gram }
    \State $\mathcal{D}_{CS} \gets \emptyset$
    \For{$text_i$ $\in \mathcal{T}_{CS}$} 
    \State $W$ $\gets$ \textsc{getConsecUnits}($text_i$)
        \For{$w_i \in$ $W$ }
            \State $cut \gets$ \textsc{SampleUnit}$(\mathcal{D}[w_{i}])$
            \State $cuts_i \gets$ \textsc{overlapAdd}($cuts_i$,$cut$)
            \State $cuts_i$ $\gets$ \textsc{NormalizeEnergy}($cuts_i$)
            \State $cuts_i$.text $\gets$ $cuts_i$.text + $w_{i}$ \Comment{Append  $w_{i}$}
        \EndFor
        \State $\mathcal{D}_{CS}$.append($cuts_i$)
    \EndFor
    \State \Return $\mathcal{D}_{CS}$
\EndFunction

\end{algorithmic}
\end{algorithm}
Using the alignments from monolingual data and maximum n-gram size, \textsc{SetupSupervisions($\cdot $)} creates a collection $\mathcal{D}$ of audio segments corresponding to each n-gram unit. Consecutive n-gram units are matched from alignments, starting with $n$ and progressing to unigrams. If an n-gram is absent, the algorithm backs off to an ($n-1$) unit. In \textsc{GenerateCollage($\cdot $)}, the function \textsc{getConsecUnits($\cdot $)} returns all consecutive ($1:n$) units. Each n-gram unit is randomly drawn from its respective collection using \textsc{SampleUnit($\cdot $)}. These segments are appended to the current spliced utterance with \textsc{overlapAdd($\cdot $)} described in \S\ref{subsec:overlap-add}, and the resulting combined utterance undergoes energy normalization \textsc{NormalizeEnergy($\cdot $)} from Eq\ref{eq1}.



\subsection{Zero-shot CS framework}
\label{subsec:zero-shot}
In this case study we focus on generating Arabic-English code-switching (CS) data, operating under the assumption that no Arabic-English CS training data is available. To generate speech data using the Speech Collage method, we require CS text. We generate the CS text from monolingual resources using the lexicon-based (Random) replacements approach described in \cite{hussein2022code}. The approach entails the following steps:
\begin{enumerate}
    \item \textbf{Parallel Text Translation}: We leverage a public Arabic-English Machine Translation System\footnote{API access available from \url{https://mt.qcri.org/api}} to generate the parallel English text from the Arabic transcription.
    \item \textbf{Word Level Alignments}: After translation, we fine-tune multilingual BERT (mBERT)  \cite{dou2021word} to obtain the word-level alignments.
    \item \textbf{Random Replacement}: Given the alignments, Arabic words are randomly substituted with their corresponding English words at a rate of $20\%$, as suggested by \cite{hussein2022code}.
\end{enumerate}

\subsection{End-to-End Speech Recognition}\label{AM}
In this work, we utilized the end-to-end (E2E) ASR conformer architecture \cite{conformer}, with the ESPNET toolkit \cite{guo2020recent}. The E2E-ASR implementation consists of a conformer encoder and a transformer decoder. Both are multiblocked self-attention architectures with the encoder further enhanced by an additional convolution module. The ASR task is formulated as Bayesian decision finding the most probable target word sequence $\hat{\mathbf{Y}}$, from all possible outputs $\mathbf{Y}^*$, by selecting the sequence which maximizes the posterior likelihood $P(\mathbf{Y}|\mathbf{X})$, given T-length sequence of D-dimensional speech features, $\mathbf{X} = \{\mathbf{x_t} \in \mathbb{R}^D| t = 1, \cdots, T \}$.
For text tokenization, we used word-piece byte-pair-encoding \cite{kudo2018sentencepiece}. The total loss function $\mathcal{L}_{\mathrm{asr}}$ is a multi-task learning objective that combines the decoder cross-entropy (CE) loss $\mathcal{L}_{\mathrm{ce}}$ and the CTC loss \cite{graves2006connectionist} $\mathcal{L}_{\mathrm{ctc}}$.
\begin{equation}\label{eq6}
\mathcal{L}_{\mathrm{asr}}=\alpha \mathcal{L}_{\mathrm{ctc}}+(1-\alpha) \mathcal{L}_{\mathrm{ce}}
\end{equation}
where $\alpha$ is used for interpolation. In our approach, the conformer is initially pre-trained on monolingual data and subsequently fine-tuned on  monolingual and synthetic CS speech combined.  

\subsection{Code-Mixing Index}
To quantify the amount of code-switching we use \textit{Code-Mixing Index} (CMI) metric \cite{gamback2016comparing}. The CMI for an utterance is defined as:
\begin{equation}
CMI=\frac{\frac{1}{2} *\left(N-\max_{i} \right)+\frac{1}{2} P(x)}{N}
\end{equation}
Where $max_{i}$ represents the number of words in the dominant language $i$, $N$ is the total word count in the utterance, $P$ is the number of code alternation points, with the constraint $0 \leq P<N $. A low CMI score indicates monolingualism in the text whereas the high CMI score implies high degree of code-mixing in the text.

\section{Data and Experimental Setup}
\label{sec:typestyle}

\textbf{In-domain:} The target domain we are considering is the Mandarin-English code-switching, specifically SEAME \cite{lyu2010seame}. In this scenario, we utilize monolingual training data from Chinese AISHELL-1 \cite{bu2017aishell}, $100$h of English data randomly sampled from Tedlium3 \cite{hernandez2018ted} and SEAME text \cite{lyu2010seame} to generate $62.2$ hours of CS data. Evaluation is performed on SEAME test sets (devman and devsge), measuring mixed error-rate (MER) that considers word-level English and character-level Mandarin. We also report WER on monolingual English and CER on monolingual Chinese subsets.\\
\textbf{Zero-shot:} 
For this scenario, we use monolingual training data from MGB-2 \cite{ali2016mgb} and Tedlium3. We generate $80$ hours of CS data using synthetic CS text described in \S\ref{subsec:zero-shot}. Evaluation is conducted on ESCWA \cite{ali2021arabic}, which is a real Arabic-English CS dataset.  \\
\textbf{Data pre-processing:} All audios are augmented with speed perturbations ($0.9$, $1.0$ and $1.1$) and transformed into 83-dimensional feature frames ($80$ log-mel filterbank coefficients plus 3 pitch features). Additionally, we augment the features with \texttt{specaugment}, with mask parameters $(mT,mF,T,F)=(5,2,27,0.05)$ and bi-cubic time-warping.\\
\textbf{Models:} the conformer encoder consists of $12$ blocks, each with 2048 feed-forward dimensions, $256$ attention dimensions, and $4$ attention heads. The transformer decoder has $6$ blocks with configurations similar to the encoder. We combine $2622$ Mandarin characters with $3000$ English BPE units for \textbf{In domain} scenario. As for the \textbf{Zero-shot} scenario we  use a  shared Arabic-English vocabulary of size $5000$ BPE.
Our training configuration utilizes Adam optimizer with a learning rate of $0.001$, warmup-steps of $25$K, a dropout-rate of $0.1$ and $40$ epochs. We use joint training with hybrid CTC/attention by setting CTC weight $\alpha$, Eq \ref{eq6}, to $0.3$. During inference, we use a beam size of $10$ with no length penalty. For the language model (LM), we train a long short term memory (LSTM) with $4$ layers, each of $2048$ dimensions, over $20$ epochs. When integrating LM with E2E-ASR, we apply an LM weight of $0.2$.   

\section{Results and Analysis}

\subsection{In-domain CS text}
\label{sec:in-domain}
We examine the impact of augmenting data with generated CS speech from monolingual data, particularly by integrating in-domain CS text. The results, presented in Table \ref{tab:in-domain}, are based on the SEAME evaluation.
\begin{table}[b!]
    \centering
    
     \caption{Comparison of the CER/WER/MER results on SEAME. \textbf{CS}: generated CS using in-domain SEAME text. \textbf{Mono}: baseline trained on monolingual data, \textbf{(Unigram, Bigram)}: generated CS using (unigram, bigram) units, \textbf{SE}: signal enhancement from \S\ref{sec:speech-collage}, \textbf{SEAME-ASR}: topline model trained on SEAME.}
    \resizebox{0.48\textwidth}{!}{
\begin{tabular}{l c c c c c c}
\toprule
\multirow{2}{*}{\textbf{Model}} & \multicolumn{3}{c}{\textbf{DevMan}} & \multicolumn{3}{c}{\textbf{DevSge}} \\
\cmidrule(lr){2-4} \cmidrule(lr){5-7}
& CER-MAN & WER-EN & MER & CER-MAN & WER-EN & MER \\
\midrule
Mono & 37.2 & 67.4 & 32.9 & 56.7 & 47.5 & 38.4 \\
~+ SEAME-LM & 36.4& 65.9& 32.2 & 55.2 & 46.5 & 37.6 \\
~+ CS-Unigram & 31.5 & 55.3   &28.4 & 47.5 & 42.2 & 34.4 \\
~+ CS-Unigram-SE & 29.7 & 53.7 & 27.2& 44.0 & 40.9 &33.0 \\
 ~+ CS-Bigram-SE &\textbf{27.2} & \textbf{47.9}& \textbf{25.4} & \textbf{39.7} & \textbf{38.1} & \textbf{31.4}\\ \midrule
 SEAME-ASR (topline) & 15.1& 28.8 & 16.5 &21.7 & 28.7 & 23.5\\
\bottomrule
\end{tabular}
}

    
    \label{tab:in-domain}
\end{table}
The results from \textit{Mono}, obtained by training on monolingual Chinese and English data, act as our baseline. A shallow fusion with a \textit{SEAME-LM}, trained on SEAME text data, results in a marginal relative reduction: up to $2$\% in MER. However, simple CS augmentation using unigram units yields up to $15.3$\% relative reductions in MER, compared to \textit{Mono}. By further enhancing the audio quality of the generated data, we achieve an overall relative improvement of up to $34.4$\% in MER compared to the \textit{Mono}. Finally comparing our best results to ASR trained on SEAME, the absolute gap is up to $8.9$\% MER. Given that we utilize SEAME text for data generation, this gap can be attributed to audio mismatches. Thus, we anticipate that further enhancements in audio quality to align with SEAME will bridge this gap.

\subsection{Zero-shot CS}
\label{subsec:zero}
We investigate the effects of augmenting the dataset with CS speech, generated from monolingual data and synthetic CS text. This synthetic CS text is produced from the monolingual Arabic MGB-2 and English Tedlium3 datasets, as described in \S\ref{subsec:zero-shot}. Our evaluations, detailed in Table \ref{tab:in-domain}, utilize the ESCWA dataset. Operating under our assumption that we do not have access to real CS data, we use the merged evaluation sets from MGB-2 and Tedlium3 to select the best epochs for the model.
The observations align with those from \S\ref{sec:in-domain}: the \textit{CS-Unigram} yields relative reductions of $12.3$\% in WER and $22.8$\% in CER.
\begin{table}[bt!]
    \centering
     \caption{Comparison of the CER/WER results on ESCWA. \textbf{CS}: data generated using synthetic CS text. \textbf{Mono}: baseline trained on monolingual data, \textbf{(Unigram, Bigram)}: generated CS using (unigram, bigram) units, \textbf{SE}: signal enhancement from \S\ref{sec:speech-collage}}
    \resizebox{0.48\textwidth}{!}{
\begin{tabular}{l c c c c c c c }
\toprule
\multirow{2}{*}{\textbf{Model}} & \multicolumn{2}{c}{\textbf{MGB-2}} & \multicolumn{2}{c}{\textbf{TED3}}& \multicolumn{2}{c}{\textbf{ESCWA}}  \\
\cmidrule(lr){2-3} \cmidrule(lr){4-5} \cmidrule(lr){6-7}
& CER & WER & CER & WER &CER & WER\\
\midrule
Mono & \textbf{6.1}& \textbf{12.9}& \textbf{4.4}&\textbf{8.5} & 31.1 & 48.7  \\
~+ CS-LM & 6.3 & 12.5 & 4.6 & 8.7 & 38.0 & 57.0  \\
~+ CS-Unigram &6.9 & 14.6& 5.2 & 10.1 & 24.0 & 42.7  \\
~+ CS-Unigram-SE &7.0 &14.7 & 10.4 & 5.4& 23.1 & 42.0 \\
~+ CS-Bigram-SE & 7.0& 14.7& 10.2& 5.2&\textbf{22.5} & \textbf{40.8}\\
\bottomrule
\end{tabular}
}
    \vspace{-1em}
    
    \label{tab:in-domain}
\end{table}
 Interestingly, the results from shallow fusion with $Mono+CS$-$LM$ consistently underperform when compared to $Mono$. Moreover, enhancing the quality of the generated audio further improves results, leading to an overall relative improvement of $16.2$\% in WER and $27.6$\% in CER compared to \textit{Mono}. It's noteworthy that, on monolingual data, performance deteriorates with CS augmentation. This suggests model bias towards code-switching and a reduced inclination for monolingual data. We further analyze this observation in \S\ref{subsec:analysis}.

 \begin{figure}[hbt!]
\centering
\includegraphics[width=.9\columnwidth]{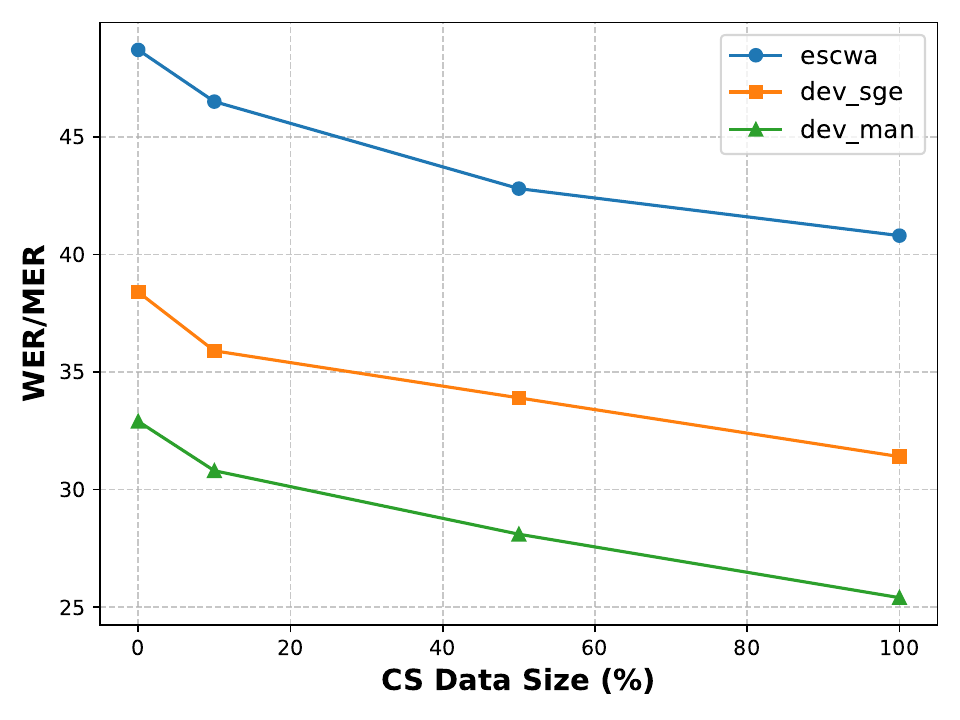}

\caption{WER/MER at different percentages of generated CS data where \textbf{0\%}: represents Monolingual, \textbf{100\%}: represents Monolingual with all generated CS.}
\label{fig:data-size} 

\end{figure}

\subsection{Generated CS data size}
We explore the impact of the amount of generated CS data size on ASR system performance.  
Figure \ref{fig:data-size} illustrates the WER at different percentages of generated CS data. In this experiment, we generated CS data with bigrams at 10\%, 50\%, and 100\%. The 0\% represents monolingual condition, while 100\% corresponds to 80 hours for Arabic-English and 62.2 hours for Chinese-English. It can be observed that there is a substantial improvement when using 10\% of generated CS data. However, as the percentage of generated CS data increases, the rate of improvement decreases. This suggests that with more data, further gains can be expected, albeit at a diminishing rate.

\subsection{Analysis}
\label{subsec:analysis}
To understand the effect of our proposed CS augmentation, we measure the average CMI.
\begin{table}[tb!]
    \centering
    
     \caption{Comparison of the average CMI. \textbf{Mono}: baseline trained on monolingual data, \textbf{SE}: Signal enhancement from \S\ref{sec:speech-collage}, \textbf{Ref}: reference, \textbf{(Uni, Bi)}: generated CS using (unigram, bigram) units.}
    \resizebox{0.48\textwidth}{!}{
\begin{tabular}{l c c c c c}
\toprule
\textbf{Dataset} & \textbf{Ref} & \textbf{Mono} & \textbf{CS-Uni} & \textbf{CS-Uni-SE} & \textbf{Bi-SE} \\
\midrule
ESCWA & 15.6 & 8.7 & 10.6 & 11.6 & 10.5 \\
SEAME & 10.4 & 3.3 & 5.4 & 6.2 & 7.3 \\
\bottomrule
\end{tabular}
}

    
    \label{tab:cmi}
\end{table}
 Notably, the conventional CMI doesn't account for the accuracy of the sentence. To address this, we select predictions that closely align with the reference using a WER with heuristic threshold set at $\leq20$\%.   
It can be observed from Table \ref{tab:cmi}, that employing CS data augmentation consistently elevates the CMI. This affirms our assumption that CS augmentation enhances the model's aptitude for code-switching.

\section{Conclusion}
\label{sec:print}
We introduced a framework that generates synthetic code-switched data from monolingual corpora. Our findings demonstrate that integrating this CS data augmentation yields substantial improvements that surpass results from training exclusively on monolingual sources or simply combining with a code-switched language model. The enhancement of the generated audio's quality further improves the performance. Additionally, in a zero-shot learning scenario, our CS augmentation is superior to solely monolingual training. Finally, we show that improvements from using CS data augmentation stem from the model's increased propensity for code-switching and a decreased bias towards monolingual input.


\section{Acknowledgements}
This work was carried out during the 2022 Jelinek Memorial Summer Workshop on Speech and Language Technologies at Johns Hopkins University, which was supported by Amazon, Kanari AI, Microsoft and Google. This work was also partially supported by NSF CCRI Grant No 2120435.

%


\vfill\pagebreak

\section{REFERENCES}
\label{sec:refs}



\printbibliography
\end{document}